\documentclass[fleqn,10pt]{olplainarticle}

\title{Multipurpose \textit{in situ} cell design for 3D X-ray imaging of electrochemical processes}

\author[1,2]{Riley J. Hultquist}
\author[1,2]{David Simonne}
\author[1,3]{Sayantan Mondal}
\author[1,2,4*]{Ericmoore Jossou}
\affil[1]{Materials in Extreme Environment Laboratory, Massachusetts Institute of Technology (MIT), 77 Massachusetts Ave., Cambridge, 02139, MA,USA}
\affil[2]{Department of Nuclear Science and Engineering, MIT Cambridge, MA, USA}
\affil[3]{Department of Materials Science and Engineering, MIT Cambridge, MA, USA}
\affil[4]{Department of Electrical Engineering and Computer Science, MIT Cambridge,  MA, USA}
\affil[*]{Corresponding author: \href{ejossou@mit.edu}{ejossou@mit.edu}}


\begin{abstract}
We present the design of a modular multipurpose cell for monitoring the degradation of materials in extreme environments.
This cell decouples the reference electrode from the working and counter electrodes, permitting precise electrochemical control and measurement reliability.
The design is compatible with \num{4}\textsuperscript{th} generation synchrotron light sources, and its emphasis on modularity facilitates adaptation to different beamlines, where there may be variations in sample stage requirements and X-ray imaging techniques.
Experimental tests with the novel design demonstrate its support of real-time corrosion and hydrogen embrittlement measurements under both Bragg Coherent Diffraction Imaging (BCDI) and Dark Field X-ray Microscopy (DFXM) configurations.

\end{abstract}

\begin{document}
\flushbottom

\maketitle
\thispagestyle{empty}

\section*{Introduction}
It is well established that pure metals are thermodynamically inclined to transform into chemically stable forms, including oxides, hydroxides, salts, and carbonates \cite{Zehra2022}, through corrosion \cite{Vernon1949,Buchanan2005,Shaw2006}. 
This process is mechanistically explained by the exchange of charge between anodic (oxidizing) sites and cathodic (reducing) sites, and it is driven thermodynamically by the minimization of Gibbs free energy \cite{Shaw2006,Was2016_ch15}. 
Corrosion is observed extensively in nature and industry, and it is responsible for the familiar sight of hydrous iron oxides on refined iron and mint patina on copper alloys. 
Some instances of corrosion, like the aforementioned cases, are often benign, while others cause extensive structural damage, especially in materials subjected to harsh operational conditions. 
One of the more extreme examples can be found in the core of a thermal nuclear reactor and the structural materials that keep it operating safely and efficiently.
During operation in this environment, materials contend with corrosion in the presence of high temperatures, hydrogen, stress, and irradiation \cite{Feron2012}. 
In a pressurized water reactor (PWR), for example, the coolant consists of boric acid (\ce{H3BO3}), which modulates reactivity via the high neutron absorption cross-section of \ce{^{10}B} \cite{Hesketh2002}, and lithium hydroxide (\ce{LiOH}), which buffers the coolant pH \cite{Zinkle2013}.
Materials are subjected to this coolant chemistry at operating temperatures between \qty{280}{\degree C} and \qty{320}{\degree C} while being pressurized to \qty{150}{\bar} \cite{Feron2012}.
In light water reactors (LWRs), which include PWRs and boiling water reactors (BWRs), \ce{H2} gas is usually added to the coolant water to lower the corrosion potential, thereby reducing the incidence of stress corrosion cracking (SCC) \cite{HE_young}, a typical failure mode. 
However, some of the \ce{H2} gas may adsorb onto metal surfaces, dissociate, and diffuse into the lattice structure, leading to hydrogen embrittlement (HE) and a critical loss of ductility \cite{HE_young}.

Irradiation from fission reactions further contributes to the corrosion and HE process, leading to Irradiation Assisted Stress Corrosion Cracking (IASCC) \cite{Scott1994}. 
The fundamental understanding by which IASCC occurs is limited or, at best, conflicting \cite{Was2012}. 
This is because coupled irradiation, stress, and corrosion have historically been investigated sequentially or \textit{post hoc}: much of the current understanding of IASCC comes from assessment of reactor components outside of \textit{operando} conditions \cite{Scott1994}, making it difficult to elucidate the onset of failure.
Furthermore, most investigations have focused on correlations between environmental factors, component properties, or material features and cracking, rather than the fundamental mechanisms governing intergranular crack initiation and growth \cite{Was2007}. 
Hence, starting in 2008, Hosemann \textit{et al.} used an \textit{in situ} corrosion cell mounted on a 3 MV Pelletron Tandem Ion Accelerator to investigate the synergistic effects of corrosion and irradiation simultaneously \cite{Hosemann2008}.
However, this study and those after it primarily rely upon \textit{post mortem} characterization for quantifying the materials evolution thereby leading to conflicting reports \cite{Schmidt2021,Raiman2017,Raiman2017a}.  

X-ray imaging of electrochemical processes, such as corrosion and HE, provides a means to significantly inform failure mode models in the earliest stages. 
Over the past \num{50} years, X-ray characterization has been extensively applied to study electrochemical processes \cite{Hashimoto1979,Kendig1993,Yang2023}. In more recent experiments, computed tomography has played a significant role in revealing the mechanisms at play in electrochemical systems, such as batteries and fuel cells \cite{HEENAN201969}. 
Meanwhile, Bragg Coherent Diffractive Imaging (BCDI) has emerged as a synchrotron-based technique that utilizes coherent X-rays to study the three-dimensional morphology and lattice strain of micro- and nanocrystalline samples \cite{5773472}, and it offers several advantages over more traditional microscopic and tomographic techniques. 
Unlike Scanning Electron Miscroscopy (SEM), which has limited capabilities for strain imaging, and Transmission Electron Microscopy (TEM), which requires thin sample preparation, BCDI offers non-destructive, 3D imaging with high spatial resolution, making it ideal for studying lattice strain and defects in crystalline materials \cite{Vicente2021}. 
Furthermore, synchrotron beamlines enable a wider range of experimental conditions, many of which are incompatible with the sample environments in SEM and TEM, which generally require samples to be under vacuum. 
A BCDI measurement is performed by measuring a 3D diffraction pattern around a Bragg reflection. 
One complete measurement is sufficient to compute the out-of-plane strain perpendicular to the related set of lattice planes.
BCDI enables the study of strain evolution and material loss in single crystals, thereby elucidating the role that extreme environments play in the degradation of reactor structural materials. 

Dark Field X-ray Microscopy (DFXM) is also a non-destructive diffraction imaging technique which enables magnified high-resolution 3D mapping of sample features such as grains and domains \cite{Simons2015}. 
In contrast to BCDI, the diffracted beam in DFXM is magnified and transformed by an objective lens, which produces an inverted 2D real space projection of the sample on the detector. 
Rotating the sample around an axis parallel to the diffraction vector while incrementally changing the sample height generates a series of 2D projections, sufficient for characterizing the 3D strain field in the region of interest.
Angular rocking out-of-plane in addition to in-plane rotation yields information about mosaicity which can be used to generate local pole figures \cite{Kutsal2019}. 
2D scans can be acquired within seconds \cite{Simons2015}, making DFXM an attractive technique for studying \textit{in situ} dynamical processes like corrosion and HE. 


\begin{figure}[b!]
    \centering
    \includegraphics[width=1\linewidth]{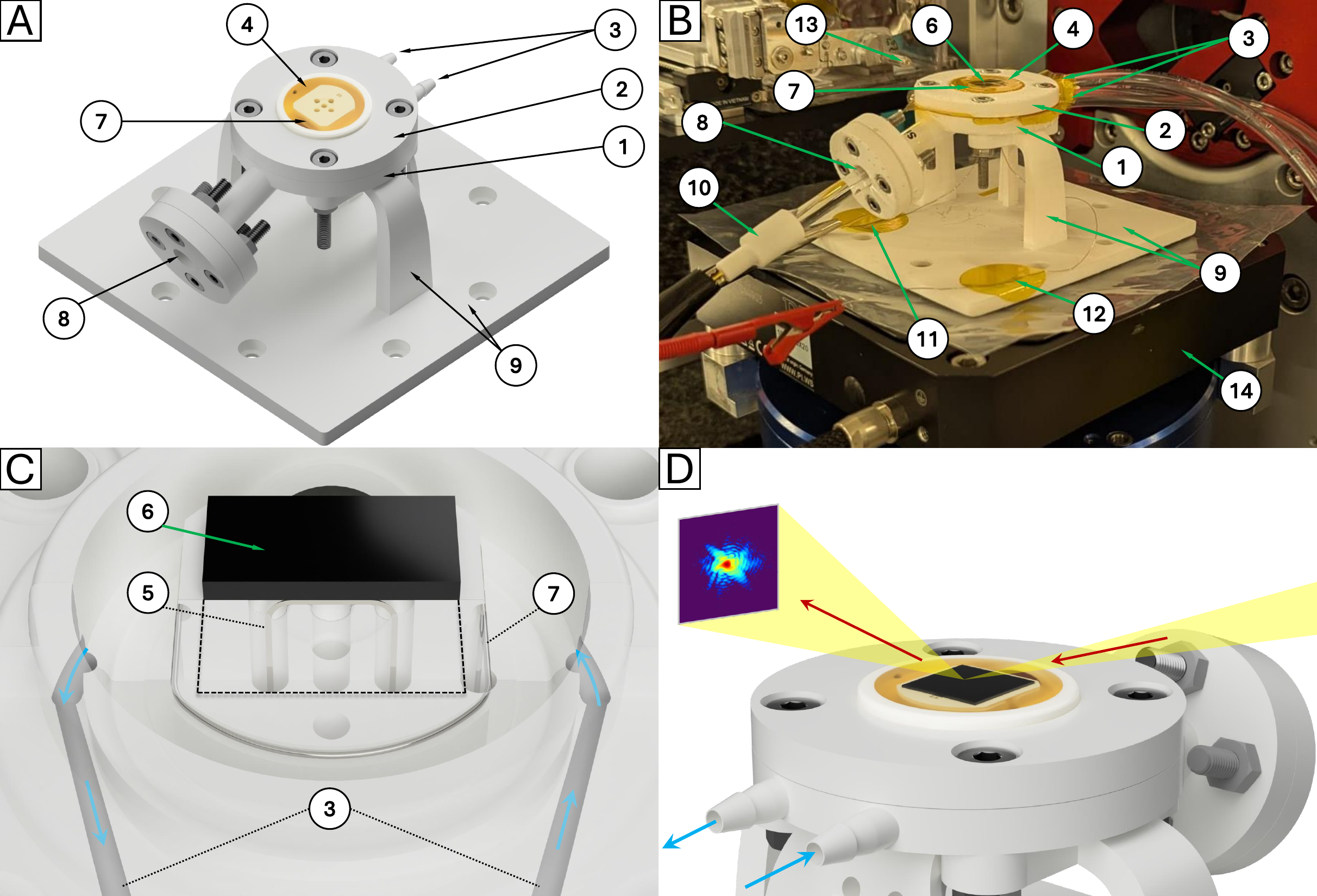}
    \caption{(A) Render of the cell with components indicated. (B) Image of the cell in use at the ID01 nanodiffraction beamline (ESRF) during \textit{operando} embrittlement. (C) Render of the sample environment with the cell lid and Kapton film removed. The front half of the cell was made transparent and the modeled sample was cut in half to show the internals. The outline of the sample contact surface is shown over the sample stage. (D) Render of the cell and drawing of the incident and outgoing X-ray beam during a BCDI experiment. A diffraction pattern is measured on the detector in the far field. Note: All numerical indicators are clarified in the text.}
    \label{fig:Cell_Components}
\end{figure}

Electrochemical experiments are vital for recognizing the role that aqueous chemistry plays in material degradation processes. 
Furthermore, irradiation-coupled corrosion and HE are of constant concern in reactor environments. 
The need for real-time, microstructural, \textit{in situ} characterization of reactor structural material degradation mechanisms motivates the incorporation of a three-electrode electrochemical cell into a compact and application-flexible design that can be used with various synchrotron techniques. 
Yang \textit{et al.} recently designed an electrochemical cell that enables the \textit{in situ} monitoring of HE \cite{Yang2025}.
In their work, a bias was applied directly to the sample (a stainless steel grain) while a Pt wire functioning as the counter electrode was placed nearby.
Their experiment corroborated real-time lattice parameter expansion as a direct result of hydrogen charging and extensively analyzed the evolution of dislocations during the process.
In this work, we describe and demonstrate the design of a cell with extended functionality which has been deployed at multiple synchrotron light sources. 
This cell enables electrochemistry experiments for microscale studies of corrosion and HE, while being chemically inert and modular for repeatability. 
It has been successfully utilized \textit{in situ} with both BCDI and `box beam’ reflection-mode DFXM, and it has been designed to be compatible with other synchrotron techniques such as Surface X-ray Diffraction (SXRD).
Corrosion and hydrogen charging data from two experiments are analyzed, demonstrating the suitability of the device for \textit{in situ} applications. 

\section*{\textit{In Situ} Cell Design}

The intended scope of the design encompasses \textit{in situ} BCDI and DFXM experiments, which have been used to demonstrate real time corrosion and hydrogen charging studies of single crystals. 
Additionally, while the cell is compact, a large field of view makes it suitable for extensive in-plane and out-of-plane coverage of the sample, enabling the use of surface characterization techniques like SXRD.

\begin{figure}[htb!]
    \centering
        \includegraphics[width=1\linewidth]{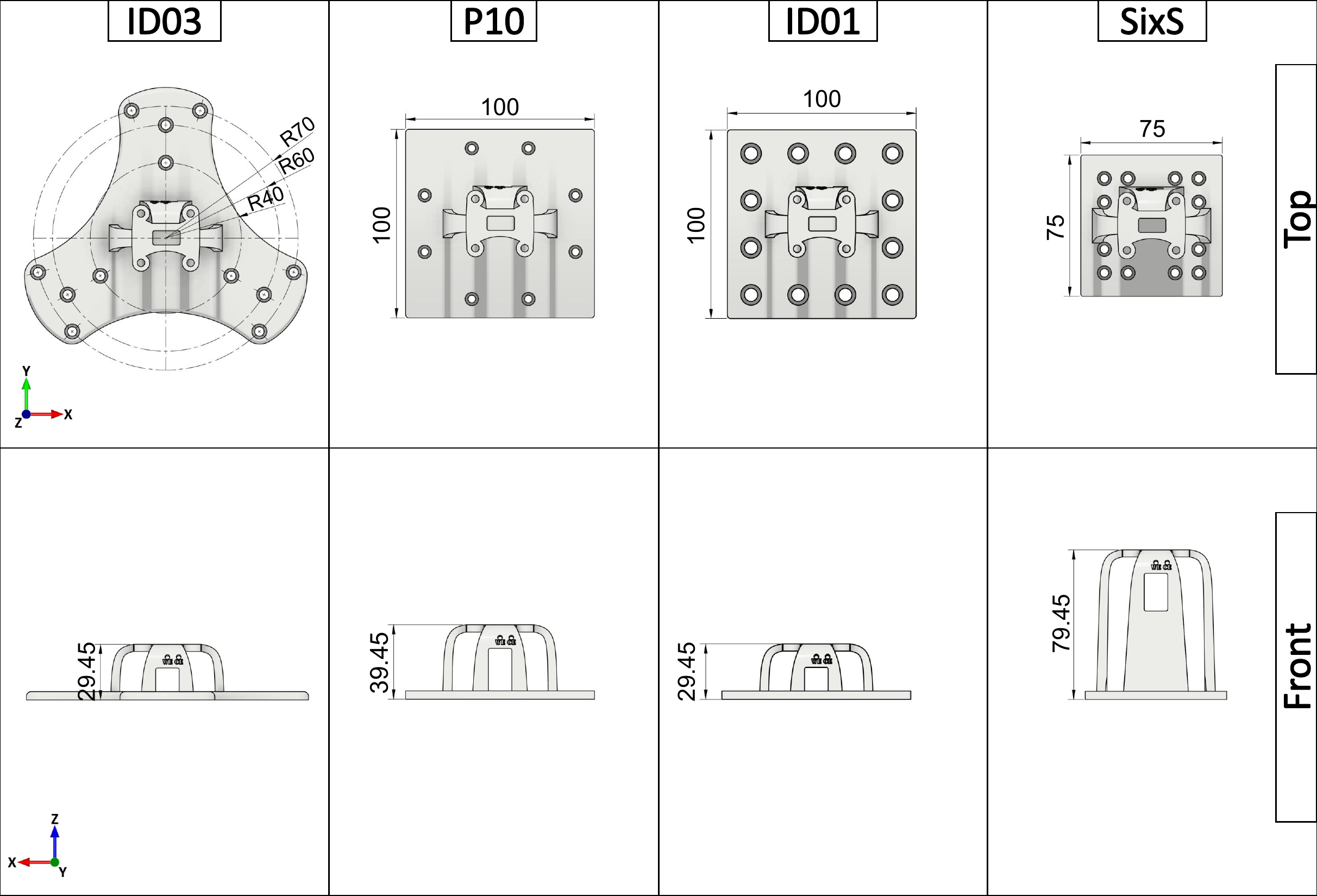}
    \caption{Drawings of the base adapters with dimensions shown in \unit{\mm}. The base adapters differ primarily in the placement of screw holes for affixing the cell to the beamline-dependent hexapod stage. Each base design is labeled with its respective beamline. The beamlines are ID03 (ESRF), P10 (DESY), ID01 (ESRF), and SixS (SOLEIL), respectively.}
    \label{fig:Base_Adapters}
\end{figure}

The cell (Fig. \ref{fig:Cell_Components}) has been designed to be compatible with multiple 3D printing platforms; however, its functionality has thus far been demonstrated using the low-force stereolithographic (LFS/SLA) Form 3+ printer developed by Formlabs \cite{Formlabs2025}. 
The cell as such consists of a strong and chemically resistant glass-filled resin that allows it to withstand sustained exposure to highly acidic and basic environments, including strong acids, such as hydrochloric acid (\ce{HCl}), and alkaline solutions, such as sodium hydroxide (\ce{NaOH}) with a pH of 10 \cite{rigid_10k}. 
It is highly resistant to several solvents, including acetone and isopropanol, indicating its compatibility with acetone-soluble epoxies that allow samples to be glued to and removed from it.
Its extremely high volume resistivity of \qty[inter-unit-product = \ensuremath{{}\cdot{}}]{1.1e15}{\ohm\cm} and surface resistance of \qty{6.9e13}{\ohm} \cite{rigid_10k} ensure that the cell body does not influence electrochemical measurements or functionality. 
Using a low-density resin keeps the total cell weight under \qty{200}{\g} for all configurations, ensuring compatibility with nanopositioning stages for nanodiffraction studies.
Attached to the cell body ((1) in Fig. \ref{fig:Base_Adapters}) are two inlet/outlet connectors (3) that fit \qty{0.125}{\inch} diameter tubes. 
The sample (6) is adhered to the stage while \qty{1.5}{\mm} diameter channels in the stage allow for the use of a working electrode wire (5) in contact with the sample. 
The height of the stage is such that the sample surface sits \qty{50}{\um} above the horizon, ensuring that no part of the cell occludes incoming or outgoing X-rays from the beamline (13) and thereby allowing complete coverage of the sample.
A Kapton film (4) is adhered to the cell lid (2) which is itself affixed to the cell body by four M\num{3.5} screws. 
For HE experiments and open circuit potential (OCP) measurements, a \qty{6}{\mm} standard reference electrode (10) is inserted into an O-ring and seated within the reference electrode housing (8). 
A lid is affixed to the housing with four M\num{3.5} screws to seal the reference electrode from the outside. 
With the reference electrode inserted, \textit{ex situ} measurements of flow with a Kamoer\textsuperscript{\textregistered} KCP PRO2 show electrolyte delivery at a maximum flow rate of \num{275} $\pm$ \qty{2}{\mL/\minute}, ensuring that reaction products are cleared from the sample environment during experiments.
A counter electrode wire (7) is threaded through the bottom of the cell body and wrapped around the sample stage, avoiding direct contact with the sample. 
For clarity, the external counter electrode and working electrode wires ((11) and (12), respectively) are indicated in Fig. \ref{fig:Cell_Components}B.
The entire configuration is fixed to the beamline-specific goniometer or hexapod stage base (14), and the base adapter (9) can be changed without redesigning the main cell, making it compatible with multiple beamlines.
This design has been successfully utilized at three different BCDI beamlines and one DFXM beamline, each with a different hexapod stage mounting base, as shown in Fig. \ref{fig:Base_Adapters}. 

\section*{X-ray Measurements}
X-ray experiments were carried out at the ID01 and ID03 beamlines of the European Synchrotron (ESRF) in Grenoble, France, the SixS beamline of the SOLEIL synchrotron in Saint-Aubin, France, and the P10 beamline of the DESY synchrotron in Hamburg, Germany. 
The first BCDI measurements, conducted at ID01, successfully demonstrated the use of the electrochemical cell under an inert gas configuration \cite{Simonne2025}.       
During these measurements, argon (\ce{Ar}) was flowed through the cell to reduce any oxidative effect of the beam. 
Gas tubes were fitted to the cell inlet and outlet without leaks.
Measurements of single Bragg reflections were taken for several particles, which revealed substrate-dependent effects on the strain fields of nickel (\ce{Ni}) microcrystals on silicon dioxide (\ce{SiO2}) and \qty{0.7}{\%} niobium-doped strontium titanate (\ce{Nb}:\ce{SrTiO3}) \cite{Simonne2025}. 
At the P10 beamline, the cell was used to study the effects of PWR chemistry during BCDI measurements. 
This demonstrated the corrosion of Ni microcrystals and confirmed the chemical stability of the cell during over \num{30} hours of electrolyte exposure. 
Additionally, hydrogen charging of copper (\ce{Cu}) microcrystals on glassy carbon was investigated for several hours, with measurements showing lattice parameter expansion.

\section*{Results and Discussion}
The results of two BCDI experiments are hereafter discussed.
Data analysis was performed using the \textit{bcdi} \cite{Carnis2019}, \textit{cdiutils} \cite{Atlan_Cdiutils_A_python}, \textit{gwaihir} \cite{Simonne2022}, and \textit{PyNX} \cite{FavreNicolin2020} python packages.
For the hydrogen charging dataset, each support was generated from the autocorrelation of the particle's electronic density.
For the corrosion dataset, the support of the first scan was generated in the same manner, while subsequent scans used the initial support without updates during phase retrieval.
Phase retrieval was performed with \num{200} iterations of  hybrid input-output (HIO) \cite{Fienup78}, \num{800} iterations of relaxed averaged alternating reflectors (RAAR) \cite{Luke2004}, and \num{200} iterations of error reduction (ER) \cite{Gerchberg1972}, while the support was updated every \num{20} iterations.
After \num{400} iterations of RAAR, a pseudo-voigt point spread function was initialized and updated each cycle to take into account partial coherence \cite{Mehta2015}. Support updates were suspended during the final \num{400} RAAR iterations and resumed thereafter. \num{50} runs of the algorithm chain were performed for each scan, while the shrink wrap threshold was altered to change the support update trajectory. 
The \num{10} best reconstructions were chosen based on the homogeneity of the amplitude distributions, then the best \num{5} were selected, orthogonalized, and decomposed into a single solution \cite{FavreNicolin2020a}.
Subsequently plotted Bragg peak cross sections, phase slices, and lattice parameter slices adhere to the CXI convention. 
During the experiments, the electrolyte was pumped into the cell at a flow rate of about \qty{60}{\mL/\minute}, corresponding to the lowest setting of the pump, as measured \textit{ex situ}.

\subsection*{Ni Corrosion in PWR Chemistry}

\begin{figure}[htb!]
    \centering
    \includegraphics[width=1\linewidth]{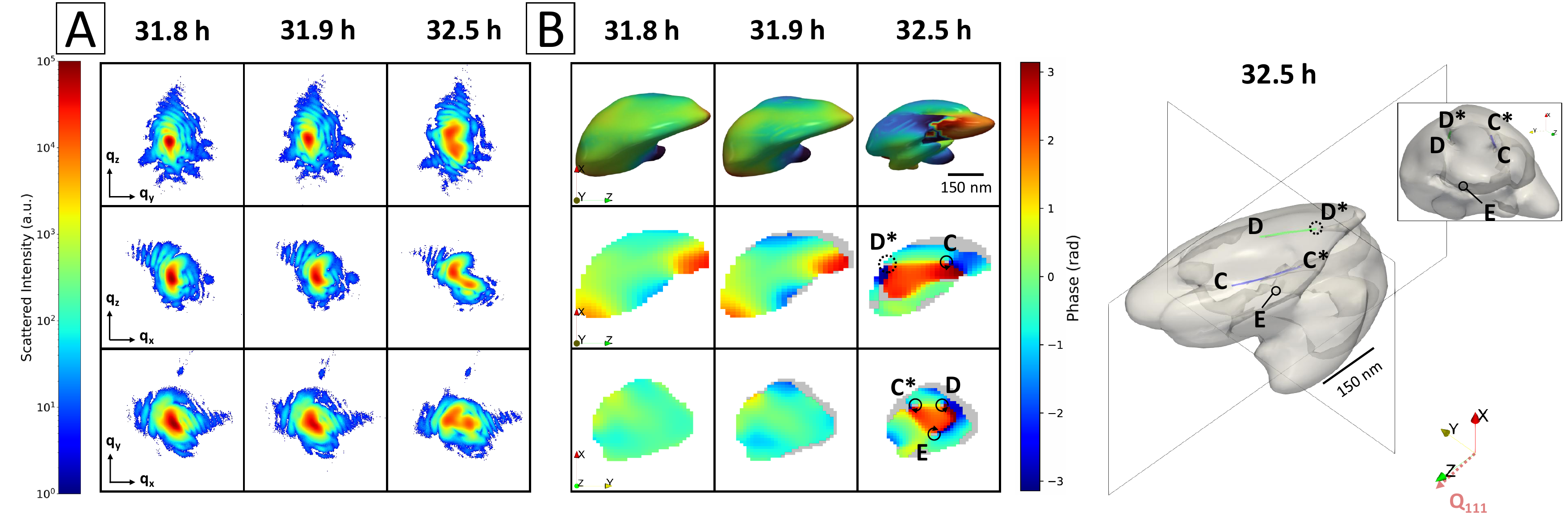}
    \captionof{figure}{Evolution of a \ce{Ni} microcrystal's Bragg peak during PWR coolant simulation. 
    (A) Center-of-mass Bragg peak cross sections are shown perpendicular to each axis. 
    (B) First row: 3D phase isosurface evolution as visualized in ParaView \cite{Ahrens2005}. Second and third rows: Phase evolution at midsection slices in the $ZX$ and $YX$ planes, respectively. Dislocations are marked C--E with arrows indicating the directions of the phase spirals. 
    An isometric view of the phase isosurface at \qty{32.5}{\hour} is shown to the right of the phase plot, as rendered in ParaView.}
    \label{fig:Corrosion}
\end{figure}

At the P10 beamline (DESY), the chemical resistance of the cell was tested during \textit{in situ} corrosion of a Ni microcrystal in PWR coolant chemistry.
The microcrystals were fabricated by solid-state dewetting of a \qty{50}{\nm}-thick Ni film deposited on an oxidized \ce{Si}(100) substrate, following an established preparation process \cite{Simonne2025} at a maximum temperature of \qty{900}{\degreeCelsius}. 
Using BCDI, the 3D (111) Bragg peak of one Ni microcrystal was measured using \qty{10.3}{\keV} X-rays by rocking the sample.
A simulated PWR coolant consisting of a solution of \ce{H3BO3} and \ce{LiOH} at a pH of 7.2 was employed to study \ce{Ni} corrosion.
The solution was calibrated by exposing the sample in the cell \textit{ex situ} for \qty{8}{\hour}, and the crystals were reviewed \textit{post hoc} with SEM to confirm their stability.
The experiment was then performed \textit{in situ}.
Therefore, $t = \qty{0}{\hour}$ refers to the start of electrolyte flow at the beamline, and the first rocking curve of this particle was acquired at $t = \qty{31.8}{\hour}$.
The flow cell was used without the reference electrode, and the entrance holes for the working and counter electrodes were sealed with epoxy to prevent electrolyte leakage during the experiment. 
Reciprocal space slices of the evolving Bragg peak are shown in Fig. \ref{fig:Corrosion}A. Additionally, the particle isosurface and phase cross sections are shown in Fig. \ref{fig:Corrosion}B.

Initially, the particle phase field appears to be continuous and differentiable. 
The Bragg peak region shows some asymmetry in the ($\vec{q}_x$,$\vec{q}_z$) and ($\vec{q}_x$,$\vec{q}_y$) reciprocal space planes, though no singularities are present in the phase. 
The second scan is largely unchanged, with roughly \qty{8.5}{\min} having elapsed after the first scan. 
The particle reconstruction suggests some shrinkage, as is evident from the exposed gray support of the first scan, which is visible behind the phase plots of the second and third scans.

The particle was scanned a third time, \qty{41}{\min} after the first scan, revealing significant changes in the Bragg peak. 
Notably, the peak appears to split bimodally, strongly implying the presence of significant structural heterogeneity in the crystal lattice. 
The large time interval without the beam between the second and third measurements reduces the likelihood that X-rays induced the observed changes.
The $ZX$ and $YX$ phase cross sections show multiple phase singularities. 
Around these singularities, the phase field forms a spiral, which is typically associated with dislocations, particularly screw-type dislocations \cite{Clark2015}. 
The singularities are labeled as C, C\text{*}, D, D\text{*} and E in Fig. \ref{fig:Corrosion}B, and the associated phase spiral directions are indicated. 
A dislocation line (blue) is drawn between C and C\text{*}, consistent with the appearance of a hollow core in the reconstructed particle isosurface displayed to the right of the phase plot. 
Another dislocation line (green) is predicted to connect D and D\text{*}, however, it is not known with certainty if the phase spiral around D\text{*} characterizes a dislocation due to being incomplete, as shown in the dashed circle in Fig. \ref{fig:Corrosion}B. 
Finally, a significant hollow core can be seen at the location of E, though a connected dislocation was not detected.
Although it is difficult to assess the nature of the dislocations conclusively with just one Bragg reflection, the reconstructed phase field clearly reflects significant defect formation over the course of the corrosion experiment.

\subsection*{Hydrogen Charging of \ce{Cu} in \ce{HCl}}

At ID01, the cell was tested to demonstrate is suitability for electrochemical hydrogen charging during \textit{operando} measurements of hydrogen-related effects on metals using BCDI.
A silver/silver chloride (\ce{Ag}/\ce{AgCl}) electrode containing saturated potassium chloride (\ce{KCl}) was introduced into the cell through the reference electrode port, and a platinum (\ce{Pt}) wire was inserted through the working electrode channel.
A separate \ce{Pt} wire was then inserted through the bottom of the cell and wrapped around the stage, acting as the counter electrode and thereby completing the three-electrode circuit. 
The sample consisted of \ce{Cu} microcrystals dewetted from a \qty{100}{\nm}-thick \ce{Cu} film deposited on a glassy carbon wafer.
Solid-state dewetting was performed at a maximum temperature of \qty{800}{\degreeCelsius} for 5 hours.
The glassy carbon substrate, purchased from MSE Supplies, has a rated resistivity of \qty[inter-unit-product = \ensuremath{{}\cdot{}}]{4.5e-5}{\ohm\cm}, making it suitable for electrochemical experiments.
The sample was exposed to \qty{0.1}{M} \ce{HCl}, while the reference electrode was used to maintain a stable charging potential of \qty{-1.0}{V} vs the standard hydrogen electrode (SHE). 
\begin{figure}[htb!]
    \centering
    \includegraphics[width=1\linewidth]{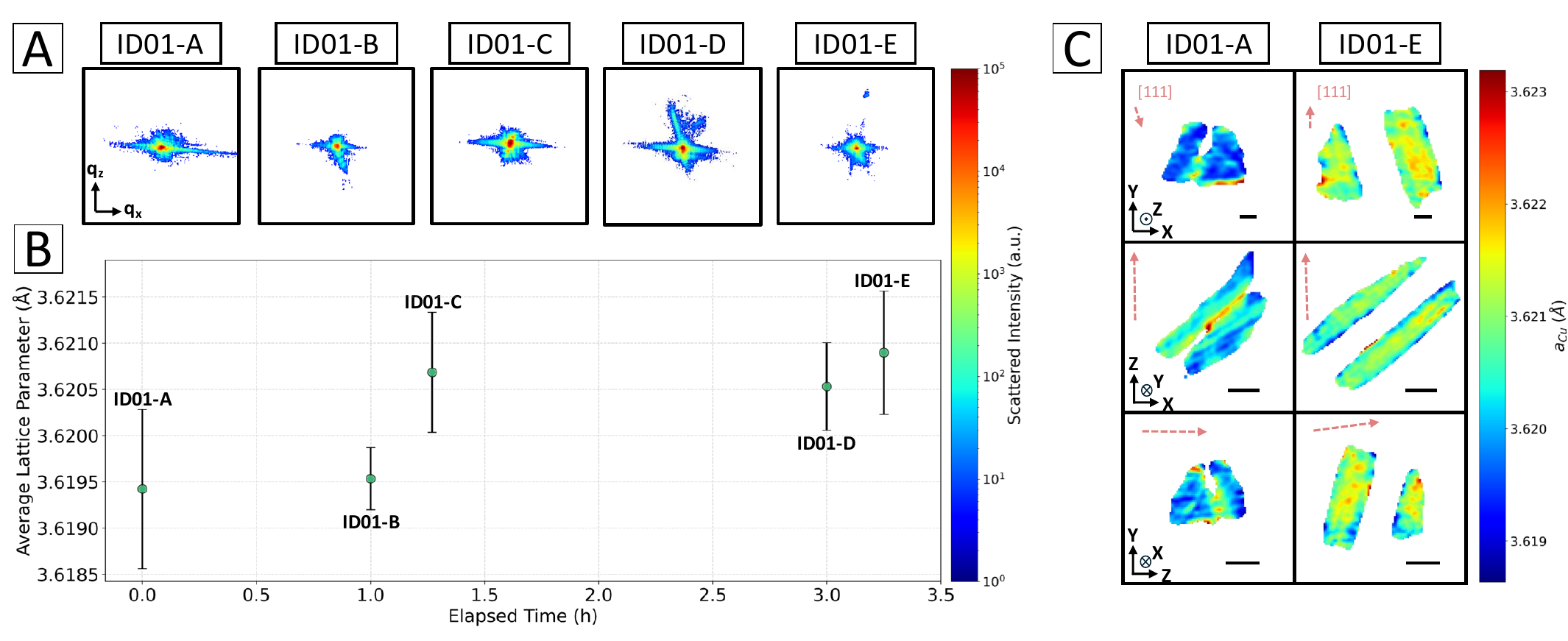}
    \captionof{figure}{Charging of Cu microcrystals on glassy carbon reveals ensemble changes in lattice parameter. (A) An $XZ$ slice of the (111) Bragg peak is shown for each particle. (B) Each particle's Bragg peak is used to calculate the homogeneous lattice parameter (Error bars represent one standard deviation of uncertainty). (C) Central slices of the lattice parameter field are shown normal to each axis for the first and last particle, along with projections of the $\hat{Q}$ unit vector. 
    Each scale bar is \qty{150}{\nm}.}
    \label{fig:Embrittlement}
\end{figure}
The introduction of a reference electrode into the cell, combined with potentiostatic regulation of the voltage, allows control of the evolution of hydrogen at the microcrystal surfaces.
Five scans were performed over the course of about \num{3.25} hours with \qty{19.9}{\keV} X-rays. 
Fig. \ref{fig:Embrittlement} shows the Cu lattice parameter evolution over the duration of \ce{HCl} exposure. 

In Fig. \ref{fig:Embrittlement}B, a trend in the average lattice parameter was derived from the measurement of each particle's lattice parameter at each timestep. 
Notably, Fig. \ref{fig:Embrittlement}C shows a fairly homogeneous expansion in the lattice of particle ID01-E at $t = \qty{3.25}{\hour}$ compared to ID01-A at $t = \qty{0}{\hour}$.  
This phenomenon is due to the relationship between \ce{H+} ions and the \ce{Cu} surfaces in the \ce{HCl} solution.
At pH 1, maintaining an electrode potential of \qty{-1.0}{V} vs SHE significantly promotes cathodic reduction of 2\ce{H+} to \ce{H2} gas at the \ce{Cu} surface.
The trend in the average lattice parameter points to the diffusion of reduced hydrogen into the \ce{Cu} lattice.
Furthermore, the formation of \ce{H2} gas bubbles was observed at the site of the reaction, providing further evidence of the HER.
The exact mechanism of lattice expansion due to the adsorbed and dissociated \ce{H2} gas is not precisely known.
Density functional theory calculations have shown that a single \ce{Cu} vacancy can trap anywhere from \num{6}--\num{8} \ce{H} atoms \cite{You2013,Du2020}.
The incorporation of \ce{H} into the \ce{Cu} bulk is expected to facilitate the formation of \ce{Cu} vacancies, occurring simultaneously with the creation of lattice strain \cite{Fotopoulos2023} and potentially leading to embrittlement.
Nonetheless, a high \ce{H} partial pressure and temperature are needed to cause significant incorporation of \ce{H} into the bulk \ce{Cu} at equilibrium \cite{Fotopoulos2023}.
Even so, cathodic reduction experiments of \ce{Cu} in \ce{H2SO4} have revealed embrittlement in \ce{Cu} (100) foils due to voids and dislocations generated by atomic \ce{H} incorporation and \ce{H2} gas bubble pressure \cite{Nakahara1989}.

\section*{Design Advancements}
\begin{figure}[htb!]
    \centering
    \includegraphics[width=1\linewidth]{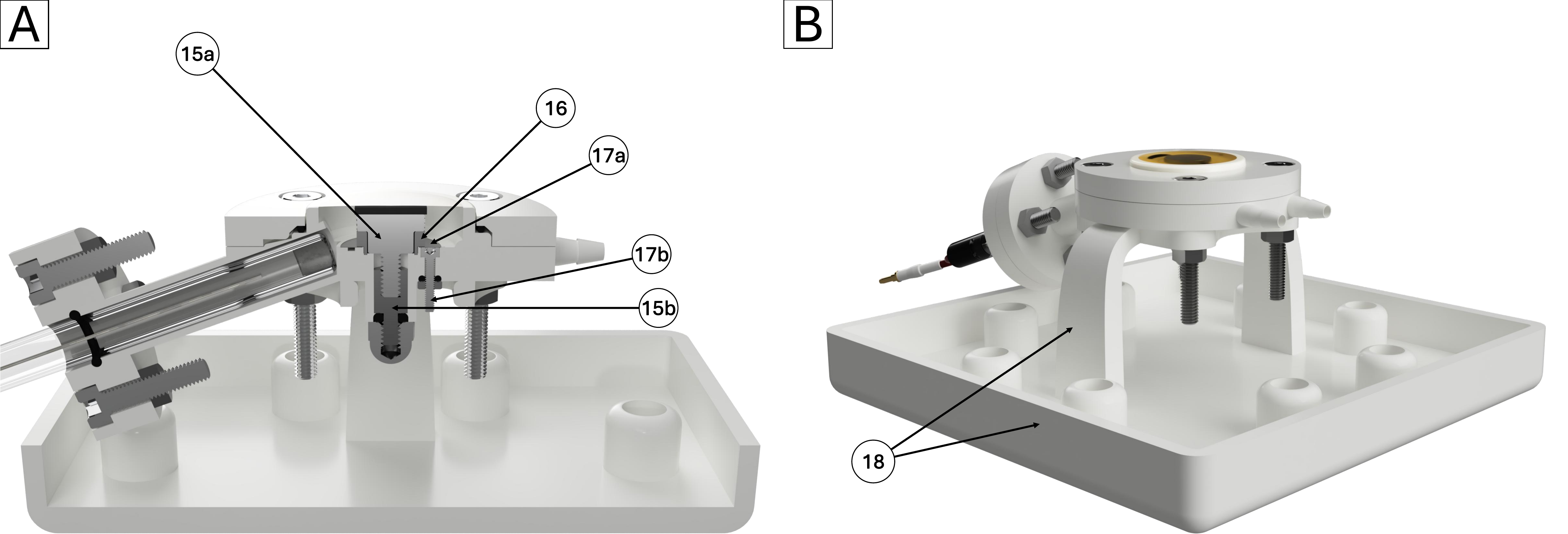}
    \captionof{figure}{(A) A cross-sectional view of the updated design with modified and additional components labeled. (B) An overview of the same design mounted on an updated, backwards compatible version of the ID01 base adapter.}
    \label{fig:Future_Design}
\end{figure}

The original design of the cell has enabled real-time monitoring of corrosion and HE under LWR-relevant coolant chemistries, with results inspiring further improvements. 
The updated version, shown in Fig. \ref{fig:Future_Design}, improves modularity and eliminates the need to print a separate cell for each sample. 
It also replaces the previous stage with an adjustable \#4--40 thread thumbscrew (15a) with a diameter of \qty{9.05}{\mm}. 
While this reduces the total surface area of the stage, it drastically enhances its functionality by providing a uniform, conductive surface on which the sample can be mounted. 
This new stage is screwed into a conductive standoff (15b) which sits fixed in the bottom of the cell. 
A set of printable spacers (16) has been modeled to fit between the thumbscrew and the cell, such that the desired sample horizon height can be achieved with different substrate thicknesses. 
The counter electrode has been replaced with a \qty{0.1}{\mm}-thick stainless steel shim (17a) which sits in direct contact with an M\num{1.4} screw. 
The counter electrode screw (17b) and the working electrode standoff (15b) can be connected externally to a potentiostat. 
The updated and prior cell designs can be mounted to updated versions of the beamline base adapters, such as the ID01 base adapter shown in Fig. \ref{fig:Future_Design} (18), which have been redesigned to contain leaks.
The replacement of the external working electrode and counter electrode nodes with threaded screws offers several advantages. 
First, screws function as reliable, conductive electrode contact points, facilitating connection to potentiostat probes while reducing the need for adhesives typically used to secure thin \ce{Pt} wires.
Second, the brittleness of thin \ce{Pt} wire makes it difficult to work into the design without snapping, while working the screws has no such associated risk.
Finally, this replacement drastically reduces the cost of fabricating the electrochemical cell, easing the burden of reproducing the tool for users and extending access to the tool for corrosion and embrittlement studies at synchrotron beamlines.

\section*{Acknowledgments}
This work was supported by the Faculty Startup Fund from the Massachusetts Institute of Technology. We acknowledge the European Synchrotron Radiation Facility (ESRF), France, for providing beamtime on beamline ID01 under proposal number 6246. We also thank DESY, Germany, for beamtime granted for the corrosion experiment on the P10 beamline under proposal number I-20240559. Surface X-ray diffraction experiments were conducted on the SixS beamline at the SOLEIL Synchrotron, France. Bragg coherent diffraction imaging (BCDI) data analysis was performed using resources of the National Energy Research Scientific Computing Center (NERSC), a DOE Office of Science User Facility supported under Contract No. DE-AC02-05CH11231 and NERSC award BES-ERCAP0029877. Sample preparation was carried out in part at MIT.nano.

\section*{Data Availability}

\sloppy The data that support the findings of this study are openly available at https://github.com/JossouResearchGroup/Electrochemical-Cell-Design.git.

\printbibliography

\end{document}